\documentclass[conference]{IEEEtran}
\IEEEoverridecommandlockouts
\usepackage{cite}
\usepackage{amsmath,amssymb,amsfonts}
\usepackage{algorithmic}
\usepackage{graphicx}
\usepackage{textcomp}
\usepackage{xcolor}
\usepackage{url}
\usepackage{tikz}

\definecolor{mygreen}{rgb}{0.0, 0.5, 0.0}

\newcommand{\ea}{\textit{et al.}}

\def\BibTeX{{\rm B\kern-.05em{\sc i\kern-.025em b}\kern-.08em
    T\kern-.1667em\lower.7ex\hbox{E}\kern-.125emX}}
\begin{document}

% FIRST SECTION
%a) the problem to be solved in the student’s research; justify why this problem is important and make clear that previous research and related work has not yet solved that problem,
%b) the research hypothesis or claim,

% SECOND SECTION
%c) the expected contributions of the research, including a discussion of related work,
%d) how the student plans to evaluate the results and to present credible evidence of the results to the community,
%e) a description of the results achieved so far, and

% THIRD SECTION
%f) the planned timeline for completion. 

\title{Unit Testing Challenges with Automated Marking}

\author{\IEEEauthorblockN{Chakkrit Tantithamthavorn, Norman Chen}
\IEEEauthorblockA{
Monash University, Australia\\
Email: \{chakkrit, norman.chen\}@monash.edu}
}

% \author{\IEEEauthorblockN{Anonymous}
% \IEEEauthorblockA{
% Anonymous University\\
% Email: Anonymous Emails}
% }

\maketitle

% This approach allows students to practice writing code that meets specific requirements while receiving immediate feedback on the correctness of their solutions. Through hands-on coding exercises, students can refine their problem-solving abilities, debug effectively, and gain real-world experience in creating robust and reliable software applications. The automated marking system not only saves instructors' time but also enables students to track their progress over time, identify areas for improvement, and engage in an active learning process that fosters a deeper comprehension of coding principles.
% the automated marking mechanism. 
% EdStem will enable students to better focus on practical and deep learning and hands-on experience in unit testing, while the automated marking mechanism allows students to understand the effectiveness of their test suites (i.e., assessing functionality completeness, correctness, and code coverage).

\begin{abstract}
Teaching software testing presents difficulties due to its abstract and conceptual nature. 
The lack of tangible outcomes and limited emphasis on hands-on experience further compound the challenge, often leading to difficulties in comprehension for students. 
This can result in waning engagement and diminishing motivation over time.
In this paper, we introduce online unit testing challenges with automated marking as a learning tool via the EdStem platform to enhance students' software testing skills and understanding of software testing concepts. 
Then, we conducted a survey to investigate the impact of the unit testing challenges with automated marking on student learning.
% We invited 330 students to participate in the survey. 
The results from 92 participants showed that our unit testing challenges have kept students more engaged and motivated, fostering deeper understanding and learning, while the automated marking mechanism enhanced students' learning progress, helping them to understand their mistakes and misconceptions quicker than traditional-style human-written manual feedback. Consequently, these results inform educators that the online unit testing challenges with automated marking improve overall student learning experience, and are an effective pedagogical practice in software testing.
\end{abstract}
% to enhance student learning experiences

% \vspace{0.5mm}


\section{Introduction}

Software testing involves various abstract concepts like test cases, test plans, coverage criteria, and different types of testing. 
% These concepts can be hard to grasp for beginners without a solid background in programming and software engineering. 
Unlike programming where students create tangible outputs, the success of software testing is often measured by what is found (i.e., defects, bugs) rather than what is created (i.e., software, apps). 
This can quickly be demotivating for some learners.
Therefore, addressing the students’ motivation and interest in software testing becomes imperative, and teaching software testing with more practical exercises is crucial in truly understanding testing processes.

However, software quality and testing subjects in many institutions often have a large number of students enrolled (e.g., 300+), which can be challenging for educators to perform effective marking and provide effective timely feedback to students.
Manual feedback can indeed be inconsistent, time-consuming, and can lead to delayed feedback. This can impact the students' learning process, leading to confusion for students who receive mixed feedback about their submissions.

To address these challenges, we introduce online unit testing challenges in order to increase students' motivation and engagement in learning software quality and testing. 
We also introduce the automated marking mechanism to effectively provide feedback to students, allowing students to understand the effectiveness of their test suites (i.e., assessing functionality completeness, correctness, and code coverage).

% TODO:ABSTRACT.

% provides anecdotal evidence with the design of courses, curricular, extracurricular activities, strategies, techniques, tools, tool adoptions, or assessment methods. An experience report should interpret the experiences in terms of actionable lessons learned.

% Experience reports will be evaluated on the basis of education and training relevance, significance, actionability, lessons, quality, and consistency of presentation.

% present an experience report

% unit testing exercises

% To address these challenges, software testing education needs to be designed with a focus on practical learning, real-world scenarios, and the integration of the latest tools and methodologies. Collaboration with industry professionals, internships, and hands-on projects can enhance students' readiness for the real-world challenges they will face in their careers as software testers.

\newpage

\section{Background and Related Work}

In this section, we provide some background and discuss the motivation with respect to related work in software testing education and automated marking in computer science.

\subsection{Software Testing Education}

Software testing education is the process of imparting knowledge and skills related to testing software applications and systems. 
It involves teaching individuals various techniques, methodologies, and best practices for ensuring the quality, reliability, and functionality of software products. Proper education in software testing is crucial to producing high-quality software that meets user requirements and performs effectively in real-world scenarios.
In particular, Lemos~\ea~\cite{lemos2018impact} found that software testing education often leads to more reliable software.

Recently, Garousi~\ea~\cite{garousi2020software} conducted a systematic mapping study of 204 papers published between 1992 and 2019. 
They found that software-testing education is becoming more active.
Most of the studies focus on proposing novel pedagogical approaches (i.e., how to teach better) and proposing tools for teaching software testing.
However, there are several challenges that are not well explored.
In particular, Garousi~\ea~\cite{garousi2020software} found that software testing is often not well accepted by students and the typical teaching practices often increase the cognitive load when learning software testing.
In addition, Aniche~\ea~\cite{aniche2019pragmatic} found that Test Coverage is one of the most difficult topics in software testing. 
They found that students commonly either miss tests, i.e., they do not provide all the expected tests for a given piece of code, or they write tests that are not totally correct, e.g., the test does not actually test the piece of code.
The findings from prior work suggest that \emph{addressing students' motivation and interest in software testing becomes imperative, and making teaching software testing more practical like real-world practices is needed.}

\begin{figure*}[t]
    \centering
    \includegraphics[width=.85\linewidth]{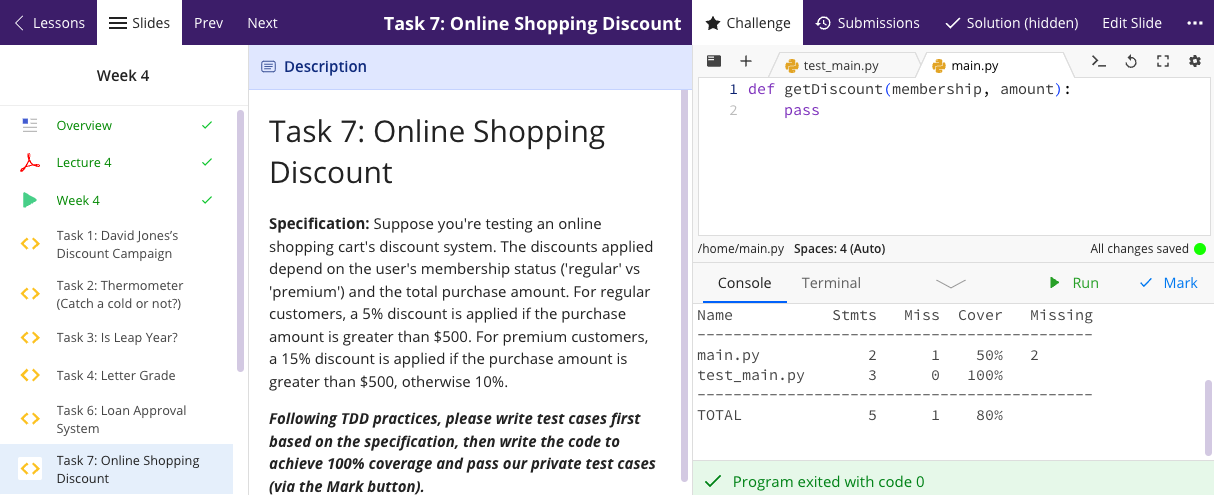}
    \caption{An Example Screenshot of Unit Testing Challenges in EdStem.}
    \label{fig:edstem}
\end{figure*}

\subsection{Automated Marking in Computer Science}

Automated marking~\cite{cheang2003automated,paiva2022automated,falkner2014increasing}, also known as automated grading or automated assessment, refers to the use of technology or software tools to evaluate and grade assignments, exams, or assessments in the field of computer science and other disciplines. This approach aims to streamline and expedite the grading process, reduce the workload on instructors, and provide timely feedback to students.

In the context of computer science, automated marking can be applied to various types of assignments, including programming projects and coding exercises. 
With various benefits, automated marking has now been used in many academic institutions.
For example, Cheang~\ea~\cite{cheang2003automated} found that an Online Judge was successfully employed in the School of Computing of the National University of Singapore for a compulsory first-year course that teaches basic programming techniques with over 700 students.
Falkner~\ea~\cite{falkner2014increasing} found that increasing marking granularity could increase the effectiveness of the automated assessment.
However, the automated marking of programming assignments (e.g., Online Judge, EdStem) only focuses on the functional completeness and correctness of the implementation or coding exercises, \emph{without evaluating the software testing aspects (e.g., how well do students test the code and how many lines of code are tested).}

% \emph{Active learning} is a teaching approach that emphasizes student engagement and participation in the learning process. It involves students actively doing something with the material they are learning, rather than just passively receiving information. 

% \emph{Constructivism} is a learning theory that suggests learners actively construct their understanding of knowledge through interactions with the environment and by integrating new information with their existing mental frameworks. 

% \emph{Experiential Learning} is a learning theory that highlights the importance of learning through direct experiences and reflection on those experiences. 

% \subsection{Subject design with Learning Theories}

% Subject design that is aligned with learning theories is crucial for creating effective and meaningful educational experiences for students. 
% Below, we briefly describe the learning theories that are incorporated into our Software Quality and Testing subject.

\section{Unit Testing Challenges with Automated Feedback via EdStem}

In this section, we present our unit testing challenges and the automated feedback mechanism.

\subsection{Online Unit Testing Challenges}

% (\emph{active learning})
% (\emph{constructivism)}

The objectives of our online unit testing challenges are to create interactive learning modules of unit testing that encourage students to actively participate in constructing their knowledge, and create unit testing challenges that are more practical and aligned with real-world scenarios.
These will enable students to apply theoretical concepts in a real-world context, thus fostering deeper understanding.

\begin{figure}[t]
    \centering
    \includegraphics[width=.85\columnwidth]{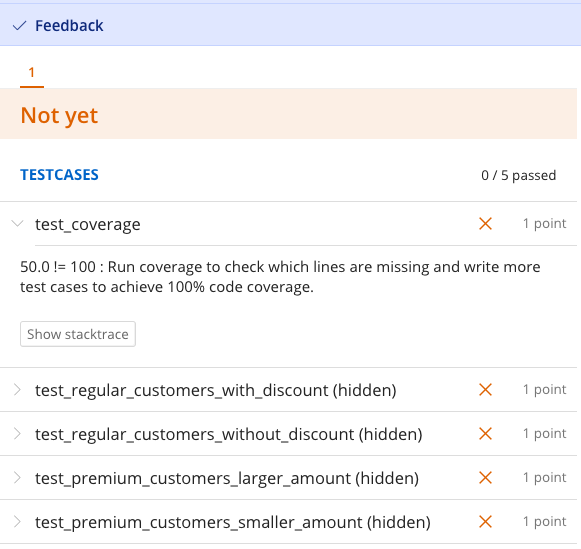}
    \caption{An example of the automated marking mechanism with feedback.}
    \label{fig:feedback}
\end{figure}

To achieve this, we use the EdStem interactive coding platform (\url{https://edstem.org/}), which offers instant programming environments for no-hassle learning.
This will ensure that students can focus on learning unit testing rather than installation issues (e.g., missing installing libraries, and different command lines among operative systems).
In particular, we use the Workspace feature in EdStem where students can execute and experiment with code and their test suite to tackle the unit testing challenges.

Figure~\ref{fig:edstem} presents an example screenshot of the unit testing challenges that we developed in EdStem.
EdStem's Code Challenge feature provides students with a way to instantly run and experiment with code with built-in support for general-purpose programming languages. 
Students will be provided with a description and have access to an embedded workspace. 
Students are encouraged to follow test-driven development (TDD) practices by writing the test cases first based on the specification, then writing the code.
In Figure~\ref{fig:edstem}, students will find each unit testing challenge task which may be completed at their own pace (i.e., \emph{self-paced learning}). 
Students will have access to a full-featured editor that they can implement their code and test files (i.e., \texttt{main.py} and \texttt{test\_main.py}) starting with the scaffold (i.e., the starting code) as seen in Figure~\ref{fig:edstem}.
Students' workspaces will be identical to the scaffold when they first open the challenge.

To self-evaluate the effectiveness of students' test cases, we configured the \texttt{Run} command to automatically execute the test cases and generate a code coverage report using the \texttt{coverage.py} library which indicates how much of the code is exercised by the students' written test cases.
In this example, the coverage report shows that there are 2 lines in the \texttt{main.py}, where 1 line is not executed (i.e., missing from the test cases), indicating that the test suite in the \texttt{test\_main.py} file only achieves a code coverage of 50\%.
Students can use the \texttt{Mark} command to submit their written code and test suites.

\subsection{Automated Marking Mechanism}

Automating certain aspects of marking can help alleviate these challenges.
EdStem's automated marking tool can execute student-submitted code against a set of private test cases to determine if the code produces the expected outputs.
However, EdStem's automated marking is generally applicable to programming tasks, i.e., only checking the functional correctness and completeness of the submitted code---\emph{without evaluating the software testing aspect (e.g., how well students test the code).}

To address this challenge, we introduced code coverage as an additional evaluation criterion into EdStem's automated marking mechanism. 
Therefore, the rubrics for the unit testing challenges will focus on the following three aspects:
\begin{itemize}
    \item \textbf{Functional Completeness:} All functions must be tested and executed via the submitted test file.
    \item \textbf{Functional Correctness:} Each of the functions must be correctly implemented according to the requirements.
    \item \textbf{Code Coverage:} All lines of code must be executed by the student's test suite (\emph{our own contribution to EdStem}).
\end{itemize}

Finally, marks for each unit testing challenge are awarded according to the percentage of the private set of test cases that
is passed by the students' code and weighted against the percentage of code coverage achieved (i.e., marks  = \% of passing test cases $\times$ \% of code coverage). 
For example, passing 90\% of the private test cases and achieving 85\% code coverage will mean a total of 0.9$\times$0.85$\times$100\% = 76.5\% marks being awarded for this task.

\begin{figure*}[t]
    \centering
    \includegraphics[width=\linewidth,clip]{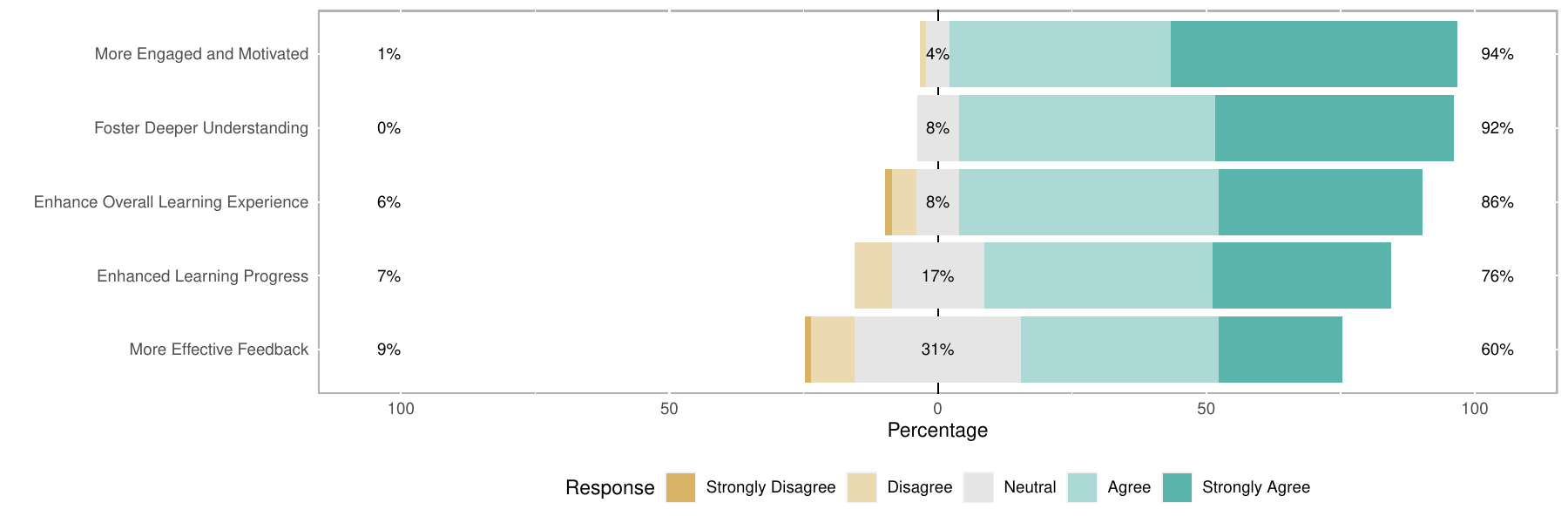}
    \caption{Survey results.}
    \label{fig:results}
\end{figure*}

\section{How do the online unit testing challenges with automated marking impact student learning?}

% \subsection{Motivation}
In this section, we present the research methodology that we used to answer this research question followed by the results.

\subsection{Designing a Subject with Learning Theories}

% \textit{Background:} 
The Software Quality and Testing subject holds a central position in the curriculum of the Bachelor of Software Engineering undergraduate degree program at our university. 
This subject is tailored for undergraduate students at Level 2 (second-year) of their studies.
This subject focuses on modern software quality assurance and testing tools and techniques to assure the quality of software systems. Students will learn different quality aspects of quality attributes, design test cases, and apply systematic testing techniques. 

\textit{Designing a Subject:} We used a responsive and iterative curriculum design approach to develop the Software Quality and Testing subject at our university in order to enhance the learning experience for students and ensure that the course content remains relevant and effective.
The subject is taught over a duration of 12 teaching weeks.
For each teaching week, the learning activities are designed as a one-hour traditional lecture-style teaching together with a two-hour studio class focused on practical hands-on experience.
In the current year of 2023, the subject has a total of 321 enrolled students. 
There are 12 studio sessions being run each teaching week, and these sessions are facilitated by two teaching associates to enhance the learning experience. 
On average, each session accommodates around 26.75 students, leading to a balanced student-to-staff ratio of approximately 13.38.
The studio sessions are used to emphasize student engagement and participation in the learning process (\emph{active learning}). 
Such learning activities will enable students to actively practise software testing with the lectures and materials they are learning, rather than just passively receiving information.
We also used EdStem to help students construct their understanding of knowledge about unit testing through interactions and hands-on practices via EdStem (\emph{constructivism}).
For each week, we run a survey to gather feedback from students as a continuous feedback loop.
This allows educators to design, develop, and improve the subject that is adaptable to the needs, interests, and abilities of the students (\emph{student-centered}).
For each week, the materials are designed to follow the journey of a software tester (e.g., starting from requirement analysis, writing acceptance testing, designing test cases, executing test cases, and defect reporting), highlighting the importance of learning through direct experiences (\emph{experiential learning}).

\textit{Designing Unit Testing Challenges.} 
Unit testing challenges are practical activities designed to help students practice and improve their skills in writing unit tests for their code. 
We integrated unit testing challenges into the weekly studio practice tasks (\textit{problem-based learning}), which could be an effective way to enhance the understanding and application of unit testing concepts among learners. 
For each teaching week that involves content on unit testing techniques, there will be 4-6 unit testing challenges for a duration of 2 hours.
These challenges typically involve writing test cases based on a specification to verify the correctness of specific functions or methods within the codebase.
Students are encouraged to follow the Test-Driven Development (TDD) approach, meaning that students should design test cases first, and then write the code in order to pass the test cases.
Feedback will be promptly provided through the automated marking mechanism.
The feedback typically includes which test cases are passed or failed, and how much code coverage is achieved.
Such prompt feedback will help students understand their knowledge gaps (e.g., incorrect implementation, missing test cases), reduce cognitive load, and better provide experiential learning opportunities.

\subsection{Survey Design}
We used an anonymous survey questionnaire to investigate the impact of unit testing challenges with automated marking on student learning, which can be used to inform educators to enhance student learning experiences in designing a software quality and testing subject.
The survey is also approved by the Human Research Ethics Committee of the University.

\textbf{Participants.} 
We invited a total of 315 enrolled students who enrolled in the subject in 2023 to participate in the survey. 
The participants are recruited by other teaching staff members (who do not belong to this research team) using EdStem announcements and during their lecture/studio classes.
The EdStem announcement includes a link to the explanatory statement stating that participation in this study is entirely voluntary and anonymous and that students may withdraw from participation at any stage with no negative repercussions from such a choice.

\textbf{Survey Questions.} Students were asked to complete surveys lasting approximately 10 minutes to gather their perceptions and the impact of online unit testing challenges with automated marking on their learning. 
We asked the following questions:

\begin{itemize}
    \item (Q1-Likert): Do you find that unit testing challenges have kept you more engaged and motivated in learning unit testing compared to traditional lecture-style teaching? 
    \item (Q2-Likert): Do you agree that the unit testing challenges foster deeper understanding and learning of the unit testing concepts and practices? 
    \item (Q3-OpenEnded): Why do you find unit testing challenges are (or are not) helpful?
    \item (Q4-Likert): Do you agree that the automated marking enhances your overall learning experience in the unit?
    \item (Q5-Likert): Do you find that our automated marking has enhanced the learning progress and helped you understand your mistakes and misconceptions quicker than traditional-style human-written manual feedback? 
    \item (Q6-Likert): Do you find that automated marking is more effective than human grading in objectively evaluating your answers (i.e., test files)?
    \item (Q7-OpenEnded): Have you encountered any challenges or concerns with automated marking? If yes, please describe briefly.
\end{itemize}

\subsection{Students' Responses}

We received a total of 113 responses, of which we excluded 21 responses due to participants not explicitly indicating their consent in the survey questionnaire.
Thus, only 92 responses will be used for the analysis, which is equivalent to a response rate of 29\% ($\frac{92}{315}$).

For the unit testing challenges, we found that:
\begin{itemize}
    \item 94\% of the participants agreed that unit testing challenges have kept them \textbf{more engaged and motivated} in learning unit testing compared to traditional lecture-style teaching.
    \item 92\% of the participants agreed that the unit testing challenges \textbf{foster deeper understanding} and learning of the unit testing concepts and practices. 
\end{itemize}

In addition, based on the analysis of the open-ended questions, we found that unit testing challenges also increase their: 
\begin{itemize}
    \item \textbf{Self-regulated learning skills}:\footnote{Self-regulated learning refers to one's ability to understand and control one's learning environment.} \textit{``It helps me understand how to code it myself without just giving me the answers.''})
    \item \textbf{Problem-solving skills}:\footnote{Problem-solving refers to one's ability to identify challenges, develop solutions, and implement strategies to overcome obstacles.} \textit{``Challenges the students to pay more attention to the small details.''}
\end{itemize}

For the automated marking mechanism, we found that:
\begin{itemize}
    \item 86\% of the participants agreed that the automated marking \textbf{enhances the overall learning experience} in the subject. For example, \textit{``Automatic feedback is good because it allows you to quickly check your answers then look for bugs if they are not passing. I think the fact that the cases are hidden is good as well as you can't just find the error from that.''}
    \item 76\% of the participants agreed that the automated marking has \textbf{enhanced the learning progress and helped them understand their mistakes and misconceptions} quicker than traditional-style human-written manual feedback.
\end{itemize}

However, only 60\% of the participants agreed that automated marking is more effective than human grading in objectively evaluating your answers (i.e., test files).
Students encountered the following challenges with automated marking:
\begin{itemize}
    \item \textbf{Suboptimal feedback} in supporting students' learning. For example, \textit{``It may be helpful to give more in-depth instructions in that case (e.g., Hint: make sure to validate input types").''} \textit{``Hidden test cases that fail should provide a hint on what is the test case about and what to do in order to pass the test case, like "Hint: What if the input type is not what you expected?)"''}
    \item \textbf{Increasing cognitive workload} when dealing with failed test cases. For example, \textit{``They are good for building general understanding but because test cases are hidden ... it creates an overall annoying and long debug process for relatively simple problems.''}
\end{itemize}

% \subsection{Lessons Learned}

Based on the challenges encountered with automated marking as described, several valuable lessons can be learned:

\begin{itemize}
    \item \textbf{Clear and Comprehensive Feedback is Crucial}: Students expressed dissatisfaction with the feedback provided by automated marking systems. To enhance the effectiveness of automated grading, it's important to offer clear and effective feedback that not only points out mistakes but also provides guidance on how to rectify them. This could involve offering more in-depth instructions, hints, and explanations tailored to the failed test cases.
    \item \textbf{Holistic Approach to Assessment}: Automated marking should not solely focus on grading but also support a holistic assessment approach. This means considering factors beyond correctness, such as code quality~\cite{iddon2023gradestyle}, problem-solving strategies, and creativity. Incorporating these elements into the grading process can provide a more comprehensive evaluation of a student's skills.
    \item \textbf{Managing Cognitive Workload}: Dealing with failed test cases can increase the cognitive workload for students, especially when debugging becomes a time-consuming process. Automated marking systems should be designed to minimize this burden. Providing step-by-step debugging guides or highlighting common mistakes can help streamline the process and prevent students from getting discouraged.

    % \item \textbf{Transparency in Test Cases}: Hidden test cases that fail without providing context or hints can be frustrating for students. To mitigate this issue, automated marking systems should make test cases more transparent. Each test case's purpose and potential pitfalls should be communicated, along with guidance on how to address them. This transparency can reduce the cognitive workload associated with debugging and enhance the learning experience.
    % \item \textbf{Balancing Difficulty and Learning}: While automated marking can be effective for building a general understanding of concepts, it's important to strike a balance between presenting challenging problems and ensuring a reasonable learning curve. Overly complex or hidden test cases can lead to frustration, hindering the learning process. Offering a mix of difficulty levels and clear instructions can help students progress without feeling overwhelmed.
\end{itemize}

\section{Conclusion \& Discussion}

In this paper, we introduce the online unit testing challenges via EdStem and the automated marking mechanism. 
Then, we conduct a survey to investigate the impact that the unit testing challenge with automated marking on student learning.
The results from 92 participants showed that the online unit testing challenges have kept the students more engaged and motivated, fostering deeper understanding and learning, while the automated feedback mechanism enhances the student's learning progress, helping them understand their mistakes and misconceptions quicker than traditional-style human-written manual feedback, which finally enhancing their overall learning experience.
These results inform educators and learning designers that the online unit testing challenges with automated marking are an effective pedagogical practice to enhance student learning experiences in software testing.

Nevertheless, students encountered two main challenges with automated marking, i.e., suboptimal feedback in supporting students' learning and the increasing cognitive workload, suggesting that clear and comprehensive feedback with a holistic approach to assessment to reduce cognitive workload.

Despite the challenges posed by online unit testing, incorporating automated marking can yield numerous benefits, particularly in the realm of large-scale teaching.
For example, 

\begin{itemize}
    \item \textbf{Motivation and Interest}: Addressing students' motivation and interest in software testing becomes more feasible with the inclusion of online unit testing challenges. The interactive and dynamic nature of automated marking systems can make the learning process engaging and stimulating. Therefore, students are likely to become more enthusiastic about mastering software testing concepts than traditional lecture-style teaching.
    \item \textbf{Timely and Actionable Feedback}: One of the most significant advantages of automated marking is the ability to provide students with prompt feedback on their assignments. This immediate feedback loop enables students to grasp their mistakes and understand the concepts they need to improve upon.
    % and implement these insights in subsequent tasks. 
    % This aspect is critical for their continuous learning and growth.
    \item \textbf{Iterability for Mastery}: Automated marking facilitates an iterative approach to learning. Students can revisit their unit testing challenges, apply the feedback they've received, and resubmit their work for marking. This iterative process allows them to refine their skills and understanding over time, promoting a deeper grasp of the subject matter.
\end{itemize}

\section{Acknowledgement}

We would like to thank Emma Yench and Paula Galvao de Barba for providing feedback on the paper, the FIT2107 2023 teaching team, and the participants.
% ., , Lil
% Zahra, Norman, Emma, all TAs

\bibliographystyle{IEEEtranS}
\bibliography{reference}

\end{document}